# Designing an Ontology for the Data Documentation Initiative

Thomas Bosch[1], Andias Wira-Alam[2] and Brigitte Mathiak[2],

[1] GESIS - Leibniz Institute for the Social Sciences, Square B2, 1,
68159 Mannheim, Germany
[2] GESIS - Leibniz Institute for the Social Sciences, Lennéstr. 30,
53113 Bonn, Germany
{Thomas.Bosch, Andias.Wira-Alam, Brigitte.Mathiak}@gesis.org

**Abstract.** An ontology of the DDI 3 data model will be designed by following the ontology engineering methodology to be evolved based on state-of-the-art methodologies. Hence DDI 3 data and metadata can be represented in form of a standard web interchange format RDF and processed by highly available RDF tools. As a consequence the DDI community has the possibility to publish and link LOD data sets to become part of the LOD cloud.

## 1  Problem

The Data Documentation Initiative (DDI) [1] in its current version 3 is an international standard for describing data from the social and behavioral sciences. Not all member of the target group have adapted it yet. In order to establish DDI as a de facto metadata standard in this field DDI should reach a broader audience.

## 2  Purpose

The goal is to motivate more people to use DDI by making it available as Linked Open Data (LOD) [2]. That way DDI data and metadata can be published and linked with other data sets in the increasingly popular LOD cloud. DDI elements can be represented by a standard based exchange format like the widely accepted and applied Resource Description Framework (RDF). The Semantic Web community can also benefit from the data delivered in DDI, allowing it to be disseminated on a large scale. This will serve both the general community and the specialists from the social sciences symbiotically.

As RDF is an established standard there is a plethora of tools which can be used to interoperate with data and metadata represented in RDF.

Use cases will be exemplified in which specific problems can't be resolved without or could be solved in a better way using the RDF representation of data and metadata specified in DDI 3.

In order to describe data and metadata specified in DDI 3 in form of RDF an ontology has to be built based on the conceptual model of DDI 3. This ontology should encompass the most relevant DDI 3 components. Further research will concentrate on expanding the developed ontology. The outline of this approach will be described. Possible applications using the RDF representation of data and metadata will be discussed to show solutions for the issues associated with the identified use cases.

## 3 Work in Progress

Figure 1 shows the planned methodology and the underlying research questions. Potential use cases are identified and will continuously be added in the research process to answer the question why an ontology of the DDI 3 data model should be built. What are the benefits associated with a DDI 3 ontology, what kinds of problems can't be solved without or can be resolved in a better way using the ontology?

An ontology engineering methodology should be specified which can be used to design an ontology based on XML Schemas. A formal methodology is needed to develop consistent ontologies, to evolve complex ontologies efficiently and for distributed ontology generation. This methodology should encompass suitable elements from state-of-the-art ontology engineering methodologies which are relevant to design ontologies on the basis of XML Schemas.

The requirements analysis is always one of the first components in existing ontology engineering methodologies. The ontology should be defined as generic as possible, not too domain-specific, modularized and compatible with other ontologies of the social science metadata standards.

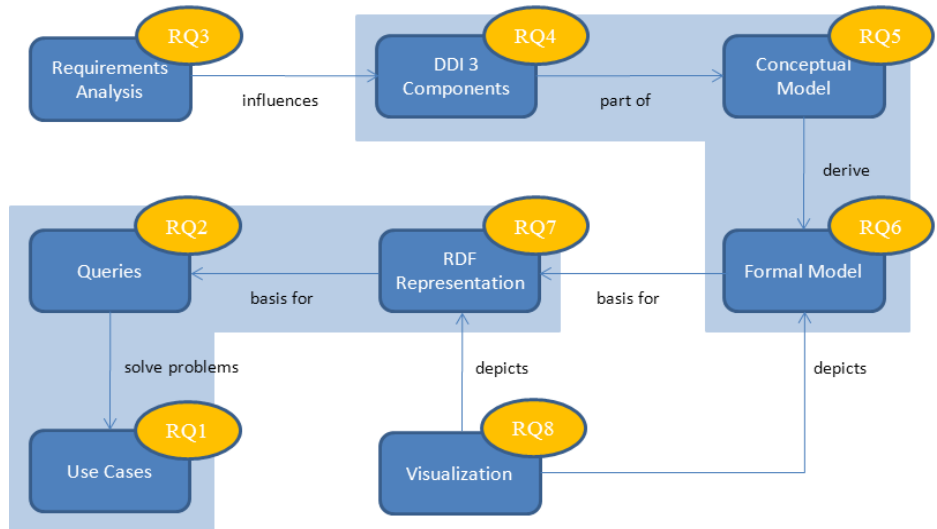

**Fig. 1.** Methodology and research questions

### 3.1 Use Cases

A number of use cases have to be detected which characterize the importance of the development of a DDI ontology.

**Semantic Queries**
What kinds of problems can be solved with the ontology to be evolved and what is the additional value? Requesting multiple distributed and merged DDI instances will be possible. To traverse the RDF graph the query language SPARQL will be applied. A SPARQL endpoint will be generated before SPARQL queries can be executed. Semantic queries will be formulated using DDI domain concepts without knowledge of DDI XML Schemas' structures.

**Publish and Link DDI Data and Metadata**
DDI data and metadata will be published in form of the standard based exchange format RDF. DDI instances can be processed by RDF tools without supporting the DDI data format and displayed using RDF browser.
After publishing public available structured data, DDI data and metadata may be linked with other data sources. Search engines for linked data can search for DDI instances which can be found in the directory of all known sources of linked data with open license (Linking Open Data Project) [2]. Crawler use RDF links between various data sources to provide expressive search functionalities. Even semantic mashups utilize linked RDF data from several data sources. The publication of linked data in the LOD cloud is the prerequisite of the development of linked data driven web applications.

**Expressiveness of Ontologies**
Ontologies based on formal logic are more expressive than XML Schemas. On that score the DDI data model can be depicted more precisely and additional concepts can be formalized as well. Ontologies can describe data models in greater detail than XML Schemas, because it does not only describe the syntax but semantics as well.

**Integration of Other Ontologies**
Domain classes of the ontology can relate to existing similar concepts of other external ontologies which are widely adopted. URIs will be used to refer to remote resources. A reasoner may use additional semantic information defined in other ontologies for deductions. As external ontologies can change over time, the referred concepts might not exist anymore. Therefore it will be necessary to jump to past versions of respective ontologies [3].

**Terminological and Assertional OWL Queries**
Both terminological and assertional OWL queries can be executed. Terminological OWL queries can be divided in checks for global consistency, class consistency, class equivalence, class disjointness, subsumption testing and ontology classification. A class is inconsistent if it is equivalent to owl:Nothing. In general this indicates a modeling error. Are there any objects satisfying the concept definition [4]? An ontology is globally consistent if it is devoid of inconsistencies. Unsatisfiability is often an indication for errors in concept definitions and for this reason you can test the quality of ontologies using global consistency checks [4]. By means of classification the ontology's concept hierarchy can be calculated on the basis of concept definitions [4].
Instance checks, class extensions, property checks and property extensions can be classified to assertional OWL queries. Instance checks are used to test if a specific individual can be assigned to a particular class [4]. The search for all individuals contained in a given class may be performed in terms of class extensions [4]. Role checks and extensions can be defined similarly with regard to pairs of individuals.

**Consistency Check of the DDI Data Model**
Verifications of class and global consistencies provide means to check the overall consistency of the DDI 3 data model and corresponding XML Schemas by association of XML Schema declaration and definitions with OWL domain concepts.

**Facilitation of Study Comparability and Enabling of Study Classification**
With the ontology to be developed and the RDF representation of DDI data and metadata the comparability of studies among different DDI instances may be facilitated. Necessary prerequisites for the study comparability have to be delineated in form of RDF. There is just a limited number of DDI 3 elements which may be compared. One possible application example would be the comparison of the general qualification for university entrance in diverse countries. Furthermore OWL reasoning will enable the classification of studies.

**Semantic References**
The meaning of references in the DDI data model will be formalized. The semantics of references to publications is not expressed in DDI 3 for instance and that's why applications can't understand the purpose of diverse references to publications in the specified data model. One possibility would be expressing these kinds of semantics with the help of RDF and Dublin Core.

**Storage of Qualitative Data**
DDI 3 in its current version 3.1 can be used to store quantitative data. Dealing with qualitative data will be implemented within the scope of the next subversion 3.2. One additional goal of the ontology creation could be to store both quantitative and qualitative data sets.

### 3.2 Ontology Engineering Methodology

An ontology engineering methodology has to be specified to design ontologies on the basis of XML Schemas. Appropriate components of state-of-the-art ontology engineering methodologies will compose this methodology. Suitable ontology engineering methodologies like the method of Uschold and King [5], the method of Grüninger and Fox [6], the METHONTOLOGY [7], Ontology Development 101 [8], On-To-Knowledge Methodology (OTKM) [9] and the Ontology Design Methodology of Stuckenschmidt [4] have been identified so far.

### 3.2 Requirements Analysis

The ontology should be defined as generic as possible and not too domain-specific like the ontology of the U.S. Census 2000 to ensure the reusability of the ontology.
The DDI 3 ontology will be divided in multiple logical modules. Existing DDI modules do not completely correspond to the data life cycle in the social and behavioral sciences. Should the ontology modules conform to the available DDI modules or to the data life cycle?
The ontology to be evolved will be compatible with other ontologies of the social science metadata standards like the SDMX (Statistical Data and Metadata Exchange) ontology. The ontology of the SDMX standard encompasses just a few elements of SDMX. Concepts of the SDMX data model will not be mapped in the new DDI ontology due to the fact that once defined concepts have to be reused in other contexts.

## 3 Future Work

If the requirements analysis is completed the conceptual model defining the structure and terminology of the domain can be specified. This continuously extended conceptual model should contain the most relevant elements of DDI 3 identified via interviews with experts of the DDI Alliance Technical Implementation Committee.

The formal model will be specified using the Web Ontology Language (OWL) 2. DDI 3 elements should be added to the ontology stepwise according to the solution of specific problems described in form of use cases. Ontology patterns, catalogue of best practices in ontology design, should be applied.

DDI 3 resources and the relationship between these resources will be represented in the RDF data format as RDF triples stored in triple stores. Both the terminological and the assertional knowledge will be visualized by means of a graphical user interface.

## 4   Conclusion

DDI will reach a broader audience by offering both data and metadata in form of the RDF format. DDI instances will be published in the LOD cloud and as a result links from and to several data sources can be followed. To reach a RDF representation of DDI in its current subversion 3.1 an ontology of the data model has to be built as a previous step. Diverse requirements have to be considered identified in the requirements analysis phase of the ontology engineering methodology. Parts of this methodology to be evolved will be appropriate components of state-of-the-art methodologies investigated in the literature. Use cases show appearing problems which will be solved using the DDI ontology.